\pdfoutput=1
\documentclass[journal,article,submit,pdftex,moreauthors]{mdpi} 

\firstpage{1} 
\pubvolume{1}
\issuenum{1}
\articlenumber{0}
\pubyear{2023}
\copyrightyear{2023}
\datereceived{ } 
\daterevised{ } 
\dateaccepted{ } 
\datepublished{ } 
\hreflink{https://doi.org/} 

\Title{An In-Silico Study on Integrated Mechanisms of Mechano-Electric Coupling in Ischaemic Arrhythmogenesis}

\TitleCitation{Title}

\Author{Viviane Timmermann $^{1, 2, 3, \dagger}$\orcidA{}, Samuel T. Wall $^{1}$, Joakim Sundnes$^{1}$,  Tarek Lawen$^{4}$, Peter A. Baumeister$^{4}$, T. Alexander Quinn$^{4,5}$, Andrew D. McCulloch$^{1,2}$, and Andrew G. Edwards $^{1,}$*}

\AuthorNames{Firstname Lastname, Firstname Lastname and Firstname Lastname}

\AuthorCitation{Lastname, F.; Lastname, F.; Lastname, F.}

\address{%
$^{1}$ \quad Department of Computational Physiology, Simula Research Laboratory, NOR \\
$^{2}$ \quad Institute for Engineering in Medicine, University of California San Diego, CA, USA \\
$^{3}$ \quad Institut für Experimentelle Kardiovaskuläre Medizin, Universitätsherzzentrum Freiburg $\cdot$ Bad Krozingen\\
$^{4}$ \quad Department of Physiology and Biophysics, Dalhousie University, Halifax, Nova Scotia, CAN \\
$^{5}$ \quad School of Biomedical Engineering, Dalhousie University, Halifax, Nova Scotia, CAN
}

\corres{Correspondence: andy@simula.no}

\firstnote{Current address: Institut für Experimentelle Kardiovaskuläre Medizin, Universitätsherzzentrum Freiburg $\cdot$  Bad Krozingen.} 

\abstract{Heterogeneous mechanical dyskinesis during acute myocardial ischaemia is thought to contribute to arrhythmogenic alterations to cardiac electrophysiology (EP). Various forms of mechano-electric coupling (MEC) mechanisms have been suggested to contribute to these changes, with two primary mechanisms being: (1) myofilament-dependent calcium release events, and (2) the activation of stretch-activated currents (SAC). In this computational investigation, we assessed the collective impact of these processes on mechanically-induced alternans that create an arrhythmogenic substrate during acute ischaemia.
To appraise the potential involvement of MEC in ischaemia-induced arrhythmias, we developed a coupled model of ventricular myocyte EP and contraction including SAC and stretch-dependent calcium buffering and release. The model, reflecting observed electrophysiological changes during ischaemia, was exposed to a series of stretch protocols that replicated both physiological and pathological mechanical conditions. Pathologically realistic myofiber stretch variations revealed calcium sensitivity changes dependent on myofilament, leading to alterations in cytosolic calcium concentrations. 
Under calcium overload conditions, these changes resulted in electrical alternans. The study implies that strain impacts cellular EP through myofilament calcium release and SAC opening in ventricular mechano-electrical models, parameterised to available data. This supports experimental evidence suggesting that both calcium-driven instability via MEC and SAC-induced effects contribute to electrical alternans in acute ischaemia.}


\begin{document}

\section{Introduction}\label{para:intro}

Coronary artery occlusion resulting in regional acute ischaemia causes 80\% of all cases of sudden cardiac death (SCD) without a prior history of heart disease \cite{Baumeister2016,Jie2010,Myerburg1997}. SCD occurs due to an acute reduction in blood supply to a region of the heart, resulting in changes in electrical, mechanical, and biochemical properties of the myocardium. During the first phase of acute ischaemia, which lasts about an hour, arrhythmogenic changes occur in the ischaemic area with marked heterogeneities at the border zone. These changes create a substrate for the induction and maintenance of lethal ventricular arrhythmia \cite{Carmeliet1999}. 
Experimental and computational studies indicate that mechanically-induced changes in cardiac electrophysiology (EP) in the ischaemic region, resulting from mechano-electric coupling (MEC), \cite{quinn2014cardiac, quinn2015cardiac, quinn2016rabbit}, may contribute both to the trigger and the substrate for ischaemic arrhythmia \cite{barrabes2002ventricular, Coronel2002, horner1994mechanically, janse2003mechanical, Jie2010, parker2004stretch, quinn2014importance}. 

The potential role of MEC in ischaemic arrhythmias has been investigated in various pig models, which have demonstrated: (1) arrhythmic episodes are less frequent in unloaded, isolated hearts than in loaded, isolated or in \textit{in vivo} hearts, with arrhythmias typically emerging from the ischaemic border (a region of tissue stretch) \cite{Coronel2002}; (2) the incidence of arrhythmias is related to the degree of dilation of the ischaemic region \cite{barrabes2002ventricular}; and (3) that MEC effects may be enhanced in ischaemic myocardium \cite{horner1994mechanically}. A subsequent computational investigation, using a mechano-electric model of the rabbit ventricles, showed that both MEC- and ischaemia-induced electrophysiological changes were necessary for ischaemia-induced arrhythmias to occur \cite{Jie2010}. The possible mechanisms responsible for these observations are presently under observation in the Langendorff-perfused isolated rabbit heart. This investigation has demonstrated that calcium buffering by BAPTA or ryanodine receptor (RyR) stabilization with dantrolene reduces the prevalence of mechanically-induced arrhythmias \cite{Baumeister2018}. Those studies indicate that the observed arrhythmias are calcium-driven events facilitated by a spatiotemporal difference between action potential duration (APD) and calcium transient duration (CaTD), as has been measured by dual voltage-calcium optical mapping (using a custom single-camera system \cite{lee2011single}). Specifically, it has been shown that ischaemia causes a rapid reduction of APD, while calcium transient (CaT) is reduced to a lesser extent and at a slower rate \cite{Baumeister2016,Baumeister2018,Wu2011}. This disparity in temporal dynamics may generate a vulnerable window in which mechanically induced alterations in cytosolic calcium levels cause the sodium-calcium exchanger (NCX) to generate electrical instabilities \cite{Baumeister2016}. 

The objective of this \textit{in-silico} study is to explore the contributions of altered myofilament calcium binding and stretch-activated currents (SAC) during mechanical dyskinesis to arrhythmias in acute regional ischaemia. We tested the influence of  MEC on ischaemic arrhythmias in two mechano-electrical myocyte models. Based on recent experimental results \cite{Baumeister2018}, we hypothesize that acute ischaemia-induced changes in myocardial contraction will lead to APD modifications. In this context, two MEC mechanisms are considered to be important: (1) activation of SAC \cite{Healy2005,quinn2017mechanically}, and (2) changes in myofilament calcium sensitivity \cite{terKeurs1998,terKeurs2006,Wakayama2001}. This study investigates the combined interplay of these two phenomena in the cases of ischaemia and heterogeneous strain to change calcium and action potential dynamics.

\section{Materials and Methods}
We developed strongly coupled mechano-electrical myocyte models to analyse the impact of stretching on APD.
We employed well-established rabbit-specific action potential (AP) models for ventricular myocytes from Shannon et al. \cite{Shannon2004} and Mahajan et al. \cite{mahajan2008rabbit} for the EP model, which integrates a realistic representation of calcium-induced calcium release (CICR) dynamics from the sarcoplasmic reticulum (SR). The Rice et al. model \cite{Rice2008} (rabbit representation for isotonic contraction) was applied to investigate myofilament mechanics. This model elucidates the crossbridge cycling and the activation of the thin filament caused by intracellular calcium binding to troponin C (TnC).

\subsection{Mechano-electrical myocyte model}\label{para:EPMmodel}
\begin{figure}[H]
\includegraphics[width=\textwidth]{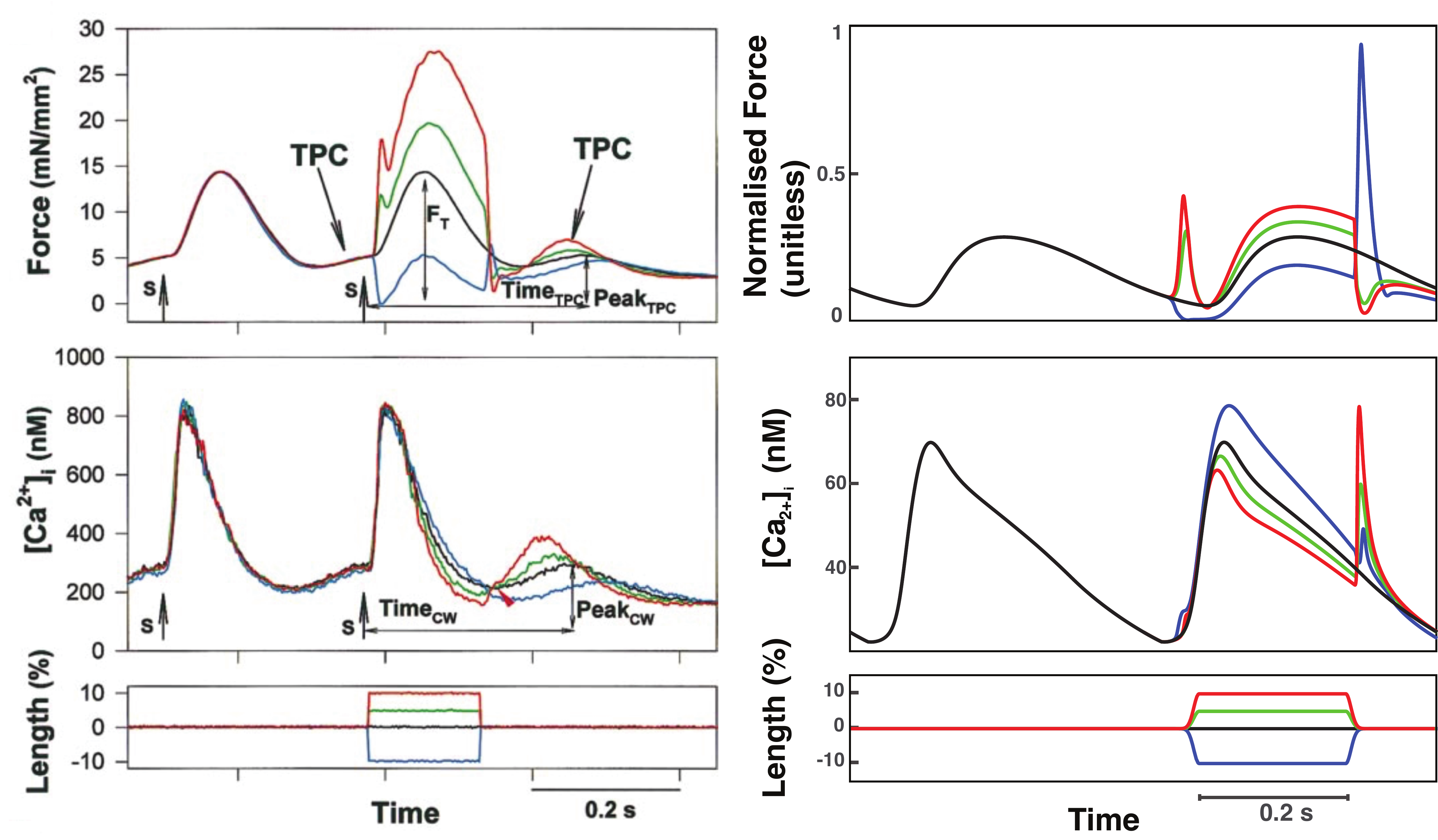}
\caption{[Left] The impact of changes in sarcomere length [bottom] on force [top] and intracellular calcium concentration [middle] during the last twitch of electrical stimulation as reported by \protect\cite{Wakayama2001}. [Right] The computational model was parameterised to reproduce the shape of the CaT amplitude, along with twitch dynamics during stretch and stretch release. The red line represents 10\% stretch, the green line 5\% stretch, the black line no stretch, and the blue line 10\% release. Similar to the experimental data, the computational model replicated an elevation in cytosolic calcium concentration due to stretch release.\label{fig:Waka}}
\end{figure}

To ensure a strong coupling, the CaT calculated in the ionic model served as input to the myofilament model, and the calcium buffering by TnC calculated in the mechanics model directly affected the ion flow of the EP model equations. For stretch-dependent changes in myofilament calcium sensitivity, we parameterised the mechanical model in order to replicate the CaT and twitch dynamics as reported in \cite{Wakayama2001} (see \cref{fig:Waka}). 

Initially, the myofilament model was modified to feature distinct on-rates for calcium buffering on the low and high-affinity sites of TnC. The two on-rates (k$_{on}^{L,H}$, respectively, displayed in \cref{table:1}) were adjusted to achieve comparable force production during isometric contraction as seen in the original Rice et al. model\protect\cite{Rice2008}, while also attaining realistic calcium buffering on TnC in the strongly coupled mechano-electrical models \cite{mahajan2008rabbit,Shannon2004}.

\begin{table}[H] 
\caption{Percentage change of the on-rates from the original model value (\protect\cite{Rice2008}) for crossbridge cycling in weakly coupled control conditions after optimizing for TnC buffering and force production in the two strongly coupled mechano-electric models.\label{table:1}}
\newcolumntype{C}{>{\centering\arraybackslash}X}
\begin{tabularx}{\textwidth}{CCC}
\toprule
\textbf{}	& \textbf{k$_{on}^H$}	& \textbf{k$_{on}^L$}\\
\midrule
\texttt{Mahajan} & -54.3\% & -87.3\%   \\
\texttt{Shannon} & +90.7\% & -10\%  \\
\bottomrule
\end{tabularx}
\end{table}

To account for strain-dependent changes in intracellular calcium handling, we modified the strain dependence of crossbridge cycling in \cite{Rice2008}. In particular, the on- and off-rates for calcium TnC buffering were modelled as an exponential function with a strain dependence as shown in \cref{eq:1}.

\begin{equation}\label{eq:1}
\begin{split}
\text{k}_{\text{on}}^{L,tot} &= \text{k}_{on}^L\cdot(e^{(\text{k}_{\text{on}}^{\lambda,L}\cdot(\lambda-1))})\\
\text{k}_{\text{on}}^{H,tot} &= \text{k}_{\text{on}}^H\cdot(e^{(\text{k}_{\text{on}}^{\lambda,H}\cdot(\lambda-1))})\\
\text{k}_{\text{off}}^{L,tot} &= \text{k}_{\text{off}}^L\cdot(e^{(\text{k}_{\text{off}}^{\lambda,L}\cdot(\lambda-1))})^{-1}\\
\text{k}_{\text{off}}^{H,tot} &= \text{k}_{\text{off}}^H\cdot(e^{(\text{k}_{\text{off}}^{\lambda,H}\cdot(\lambda-1))})^{-1}
\end{split}
\end{equation}

\begin{table}[H] 
\caption{Percentage change of the off-rates from the original model value (\protect\cite{Rice2008}) for crossbridge cycling in weakly coupled control conditions after optimizing for TnC buffering and force production in the two strongly coupled mechano-electric models.
\label{table:2}}
\newcolumntype{C}{>{\centering\arraybackslash}X}
\begin{tabularx}{\textwidth}{CCCCCC}
\toprule
\textbf{}	& \textbf{$\text{k}_{\text{off}}^{\lambda,H}$} & \textbf{$\text{k}_{\text{off}}^{\lambda,H}$} & \textbf{$\text{k}_{\text{off}}^{\lambda,H}$} & \textbf{$\text{k}_{\text{off}}^{\lambda,H}$}\\
\midrule
\texttt{Mahajan} & +30\% & 0\% & 0\% & -15\% & -30\%  \\
\texttt{Shannon} & +20\% & -10\% & 0\% & +30\% & 0\% \\
\bottomrule
\end{tabularx}
\end{table}

The experimental findings, published in \cite{Wakayama2001}, showed a decline in force when the imposed strain waveform was negative (induced shortening). Afterwards, rapid recovery to the control level and a subsequent decrease of force upon re-stretching to the reference length (blue line in \cref{fig:Waka}). However, the simulations did not elicit any force during the stretch. Nonetheless, upon subsequent re-stretching, the force of contraction increased three times beyond control. The observed discrepancy between the experimental data and simulation (\cref{fig:Waka}) observed results from differences in units. The experimental data is in total force, while the simulations use normalized force.

In addition to these calcium-mediated MEC mechanisms, we incorporated SAC in the form of a potassium selective ($I_{K0}$) and a cation non-selective ($I_{NS}$) current as developed by \cite{Healy2005} (see \cref{eq:StrainFunc}).



\begin{equation}\label{eq:StrainFunc}
\begin{split}
I_{NS} &= G_{NS} \cdot (V_m - E_{NS})\\
I_{K0} &= \frac{G_{K0}}{ 1 + e^{\frac{(19.05 - Vm)}{29.98}}}\\
I_{SAC} &= (I_{NS} + I_{K0}) 
\end{split}
\end{equation}

where $E_{NS} = -10\:mV$, $G_{K0} = 0.1\:S$, $G_{NS} = 0.08\:S$, $V_m$ represents the membrane potential in $mV$, and $x_{sarc}$ refers to the sarcomere length in $\mu m$. A previous computational study from our group investigated various conductances for both SAC for different species models during rapid stretches at distinct timings. For the present study, we chose the aforementioned SAC parameterization, which resulted in the most meaningful stretch-dependent effects on APD \cite{Timmermann2017}.

\subsection{Incorporation of ischaemia}\label{para:ischaemiamodel}
To simulate ischaemic conditions, we implemented the effects of hyperkalemia, hypoxia, and acidosis as described in \cite{Gemmell2016}. Hyperkalemia was mimicked by increasing the extracellular potassium concentration, $[K^+]_o$, and impairing the sodium-potassium pump. Hypoxia was simulated by including an adenosine triphosphate (ATP)-sensitive potassium current, $I_{K,ATP}$. Lastly, acidosis was represented by inhibiting $I_{Na}$ and $I_{CaL}$. The specific values for parameterization under acute ischaemic conditions and in the border zone are listed in \cref{tab:isch_parameters}.

\begin{table}[H] 
\caption{Values for adjusting the parameters to account for ischaemia effects in the healthy, border zone, and acute ischaemic region. $f_{NaK}$ denotes the percentage of impairment of  $I_{NaK}$, $f_{inhib}$ represents the inhibition percentage of $I_{CaL}$ and  $I_{Na}$, while $f_{inhib}$ stands for the activation of $I_{K,ATP}$. The values for acute ischaemia are obtained from \cite{Gemmell2016} and were halved to reflect the border zone.\label{tab:isch_parameters}}
\newcolumntype{C}{>{\centering\arraybackslash}X}
\begin{tabularx}{\textwidth}{CCCC}
\toprule
\textbf{}	& \textbf{Healthy}	& \textbf{Border Zone} & \textbf{Acute Ischaemia}\\
\midrule
$[K^+]_o$ (mM)  & 5.4     & 8.7         & 12.0           \\
$f_{NaK}$ (\%)  & 0       & 15          & 30             \\
$f_{K,ATP}$ (\%) & 0       & 0.4         & 0.8            \\
$f_{inhib}$ (\%) & 0       & 12.5        & 25.0          \\
\bottomrule
\end{tabularx}
\end{table}

\subsection{Altered calcium buffering on the myofilaments}
For the EP model, we used the well-established rabbit-specific AP model designed for ventricular myocytes from Shannon et al. \cite{Shannon2004}. This model depicts CICR from the SR by integrating a sub-sarcolemmal (SL) region located between the membrane and cytosolic bulk, thereby providing a more accurate description of the calcium gradients that originate during normal excitation-contraction coupling (ECC) and their interaction with the localization of calcium-sensitive transporters \cite{Edwards2017}. Furthermore, the regulation of TnC buffering in \cite{Shannon2004} describes two binding sites, high and low-affinity sites, for calcium. Both binding sites can bind calcium, but magnesium only competes with calcium for the high-affinity sites. Apart from TnC, myosin also possesses binding sites that are subject to competitive binding by calcium and magnesium.

To model the effect of stretching on the sensitivity of the myofilaments to calcium, the impact of mechanical perturbations on the binding of calcium of TnC needs to be taken into account. Any mechanical changes only alter the level of occupancy of the low-affinity TnC states in the EP model, which exerts an influence on the mechanics of both EP and sarcomere mechanics due to the strong coupling.

Alterations in membrane potential are influenced by fluctuations in currents generated by the L-type channel and NCX situated in the sub-SL. However, conventional approaches of combining mechanics and EP have presumed that the contractile proteins do not detect these gradients. Consequently, it is assumed that the myofilaments are present only in the cytosolic bulk and that modifications in the calcium sensitivity of the myofilaments affect the cytosolic calcium concentration directly. As the sub-SL is located between the cytosolic bulk and membrane, any mechanical perturbations that result in changes in CaT will not directly impact the functioning of the membrane transporters. Instead, they must first alter the calcium concentration in the sub-SL. Nevertheless, it is possible, even probable, that some of the TnC binding sites reside close to calcium release sites, allowing them to detect increased calcium concentrations during normal ECC. To investigate whether sub-SL localization of TnC can facilitate mechanically-induced alterations in calcium binding to impact sub-SL calcium concentration and the behaviour of calcium-sensitive currents, we conducted simulations. Specifically, our simulations focused on having 10\% of the myofilaments directly located within the sub-SL incorporating their calcium buffering on TnC. 

Thus, two different contractile models were computed - one with the cytosolic calcium concentration as input and the other additionally using the sub-SL calcium concentration. The resulting myocyte mechanics reflect a fractional combination of both models. For the EP model, the TnC concentration for its low- and high-affinity calcium binding sites, in addition to its high-affinity magnesium binding sites, compromised of the combined concentration fractionally from the cytosolic bulk and sub-SL. This calculation takes into consideration the differences in volume (\cref{eq:TnC_second}). 

\begin{equation}\label{eq:AffinitySites}
\begin{split}
 Mg_{Cyto} &= [Mg] \cdot fraction \\
 Mg_{SL} &= [Mg] \cdot \frac{V_{Cyto}}{V_{SL}} \cdot (1-fraction)\\
 TnC_{low_{Cyto}} &= [Ca_{low}] \cdot fraction\\
 TnC_{low_{SL}} &= [Ca_{low}] \cdot \frac{V_{Cyto}}{V_{SL}} \cdot (1-fraction)\\
 TnC_{high_{Cyto}} &=[Ca_{high}] \cdot fraction\\
 TnC_{high_{SL}} &= [Ca_{high}] \cdot \frac{V_{Cyto}}{V_{SL}} \cdot (1-fraction)
\end{split}
\end{equation}
where $[Mg] = 1$ mM, $[Ca_{low}] = 0.07$ mM, $[Ca_{high}] = 0.14$ mM, $V_{Cyto}$ volume of the cytosolic bulk, and $V_{SL}$ volume of the sub-SL.
 
This results in the following ordinary differential equation (ODE) that characterises alterations in TnC buffering for the myocyte.

\begin{equation}\label{eq:TnC_second}
\begin{split} 
 dTnC_{low_{Ca^{2+}}} &= dTnC_{low_{Ca^{2+}_{Cyto}}} \cdot fraction \\
 &+ dTnC_{low_{Ca^{2+}_{SL}}} \cdot (1-fraction)\\
 dTnC_{high_{Ca^{2+}}} &= dTnC_{high_{Ca^{2+}_{Cyto}}} \cdot fraction \\
 &+ dTnC_{high_{Ca^{2+}_{SL}}} \cdot (1-fraction)\\
 dTnC_{high_{Mg}}&= dTnC_{high_{Mg_{Cyto}}} \cdot fraction \\
 &+ dTnC_{high_{Mg_{SL}}} \cdot (1-fraction)
\end{split}
\end{equation}
where $fraction = 0.1$.

\subsection{Stretch protocol}\label{para:StretchProtocol}

\begin{figure}[H]
\includegraphics[width=\textwidth]{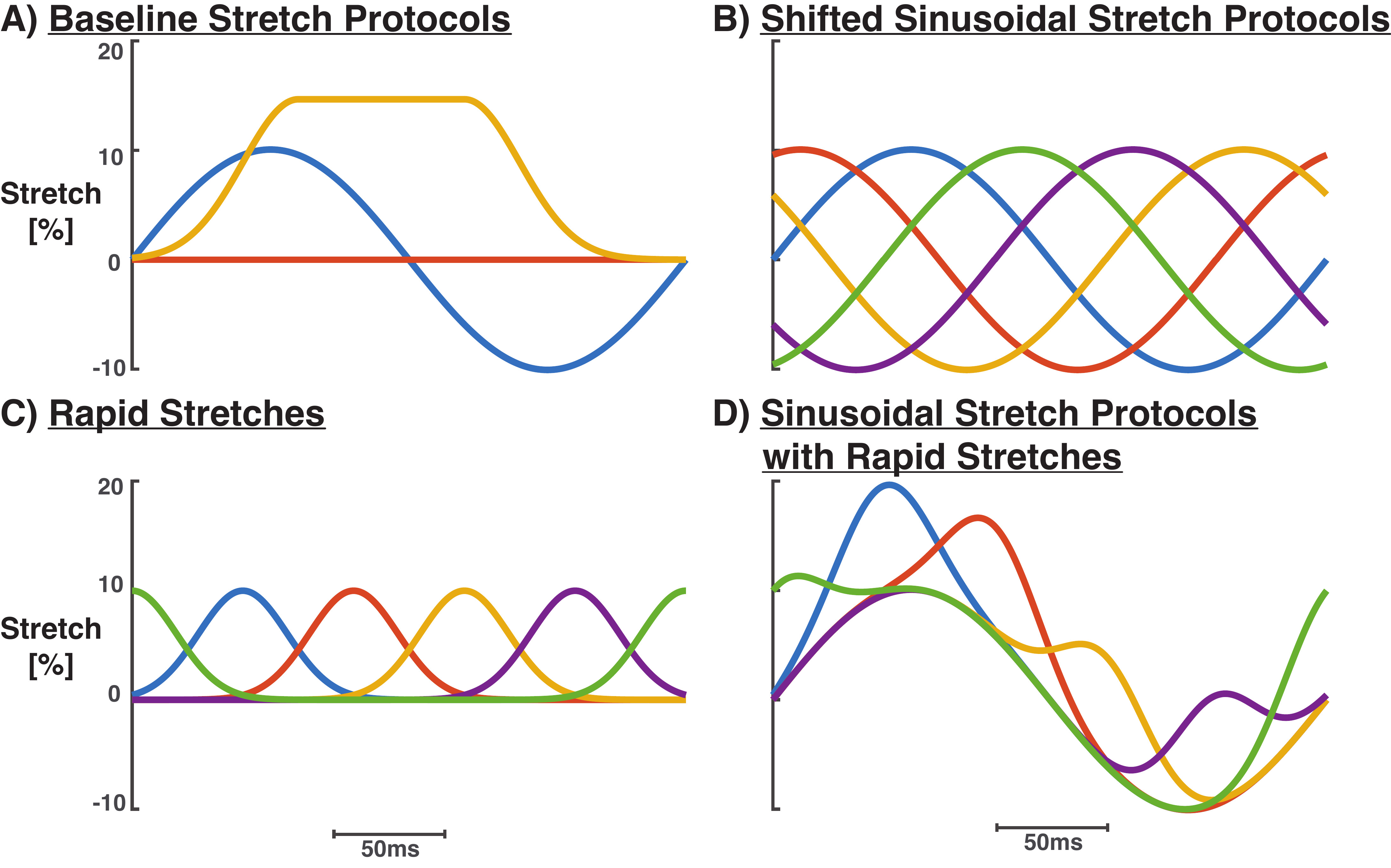}
\caption{Schematic of the various stretch protocols. Panel (A) shows three distinct baseline stretch protocols [protocol (1) in blue, protocol (2) in red, and protocol (3) in yellow]. Panel (B) illustrates an example of time-shifted protocols by 50 ms, 100 ms, 150 ms, and 200 ms, exclusively shown for protocol (1). Panel (C) showcases rapid Gaussian bell curve-shaped stretches at different timings [50 ms, 100 ms, 150 ms, 200 ms, and 250 ms]. Panel (D) exhibits an example of additional rapid stretches for the protocols exclusively shown for protocol (1).\label{figure:Figure0_second}}
\end{figure}   

Ischaemia induces regional heterogeneities in the mechanics and electrical activation, leading to irregular and non-physiological mechanical loads and anomalous synchronization of loading and electrical activation. The single-cell computational model replicates such discrepancies by conducting various stretch protocols and adjusting the magnitude, frequency, and stretch timing related to cell activation. 

To establish steady state, we performed 200 beats of pacing using a cycle length of 249.6 ms that was experimentally determined. Myocyte stretching was modelled using the 2D speckle-tracking echocardiography data in the ischaemic posterior border zone of Langendorff-perfused isolated rabbit hearts at pre-ligation (0 min), phase 1a of ischaemia (15 min), and phase 1b of ischaemia (30 min) \cite{Baumeister2018}.


We investigated three stretch protocols representing the ischaemic posterior border zone at 0 min, 15 min, and 30 min post-ligation applied to the healthy myocyte, the myocyte in the border zone, and in the ischaemic myocyte 10 mins after ligation (stretch protocol A; illustration A in \cref{figure:Figure0_second})). Additionally, these nine combinations were tested in simulations with 15\% pre-stretch (stretch protocol B). 
To introduce dyssynchrony, all stretching protocols were delayed in time by 50 ms, 100 ms, 150 ms, and 200 ms (stretch protocol C; illustration B in \cref{figure:Figure0_second})).

\section{Results}

\subsection{Strongly coupled mechano-electrics do not alter the electrophysiologic outcomes of ischaemia}
\begin{figure}[H]
\centering
\includegraphics[width=0.78\textwidth]{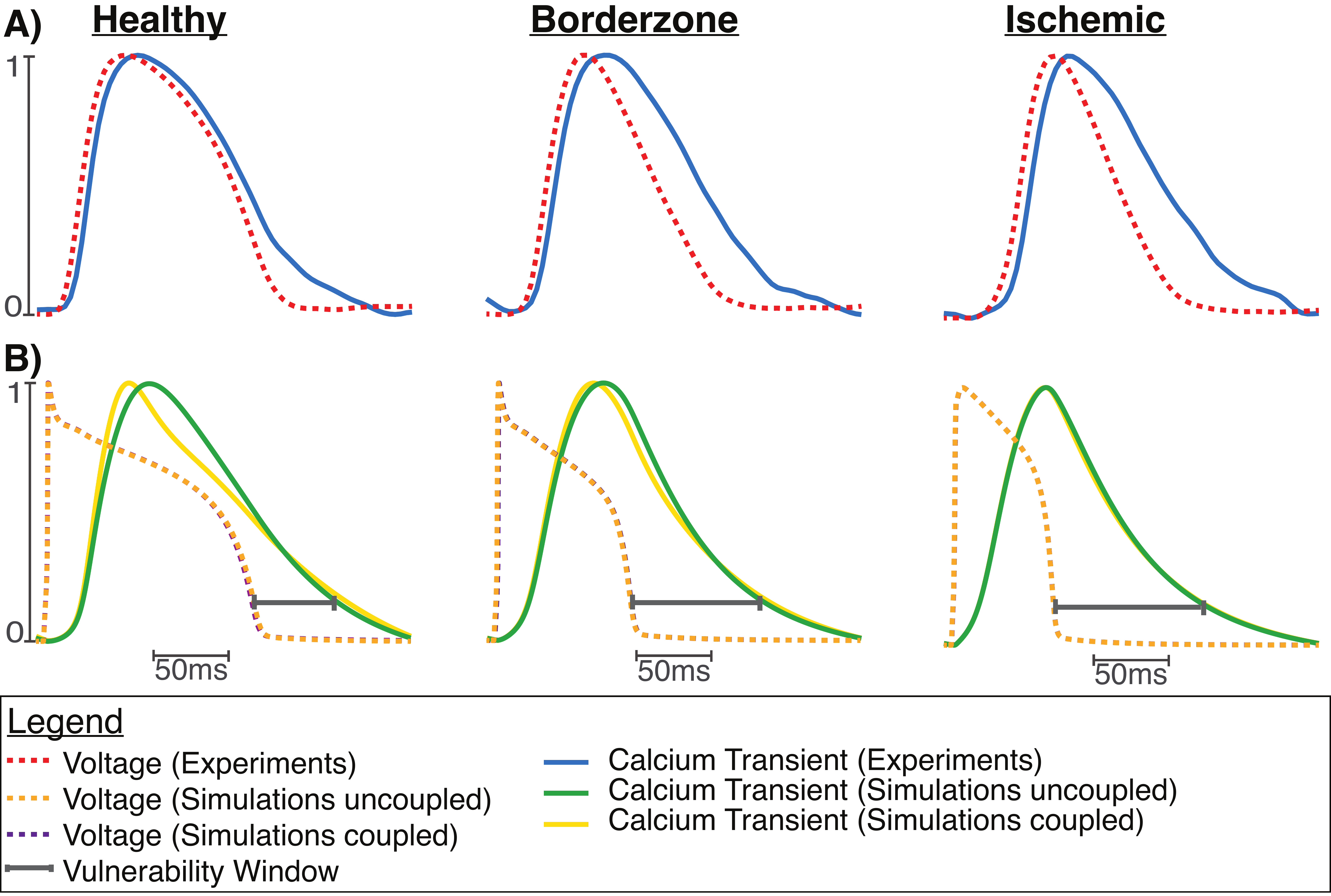}
\caption{Panel (A): Normalized CaT [solid lines] and AP [dotted lines] were measured experimentally in an excitation-contraction uncoupled Langendorff-perfused rabbit heart 30 mins post-ligation in healthy cells [left], cells located in the border zone between the healthy and ischaemic tissue [middle], and in fully ischaemic tissue [right]. Panel (B): Computational simulations of the CaT [solid lines] and AP [dotted lines] for uncoupled [yellow lines] and coupled mechano-electrics [blue lines] in a healthy myocyte [left], a myocyte in the border zone [middle], and an ischaemic myocyte [right]. The experiments and simulations indicate that with ischaemia the AP shortens, while the magnitude and time course of the CaT remains elevated. This difference in APD and CaTD creates a vulnerable window, during which NCX may generate a depolarizing current leading to early after depolarizations (EAD).\label{figure:Figure1_second}}
\end{figure}

Recent experimental evidence suggests that ischaemic arrhythmias are generated by calcium-driven and mechanics-mediated effects. Therefore, we performed simulations with varying levels of ischaemia in both weakly and strongly coupled mechano-electric systems. Our models replicated cells 10 mins after induction of ischaemia in the central ischaemic zone, the ischaemic border zone, and the normally perfused region.
During acute ischaemia, the rapid reduction of APD aligns with the experimental evidence. In contrast, CaTD decreases at a slower rate. Consequently, the disparity between APD and CaTD (as depicted by the grey bars in \cref{figure:Figure1_second}) increases towards the ischaemic region. However, the strongly coupled mechano-electric model produced almost identical CaT values to those of the weakly coupled system (as illustrated by the purple and orange lines in \cref{figure:Figure1_second}).

\subsection{Experimentally recorded stretch protocols do not alter action potential morphology}\label{section:RecordedStretch}
\begin{figure}[H]
\centering
\includegraphics[width=0.9\textwidth]{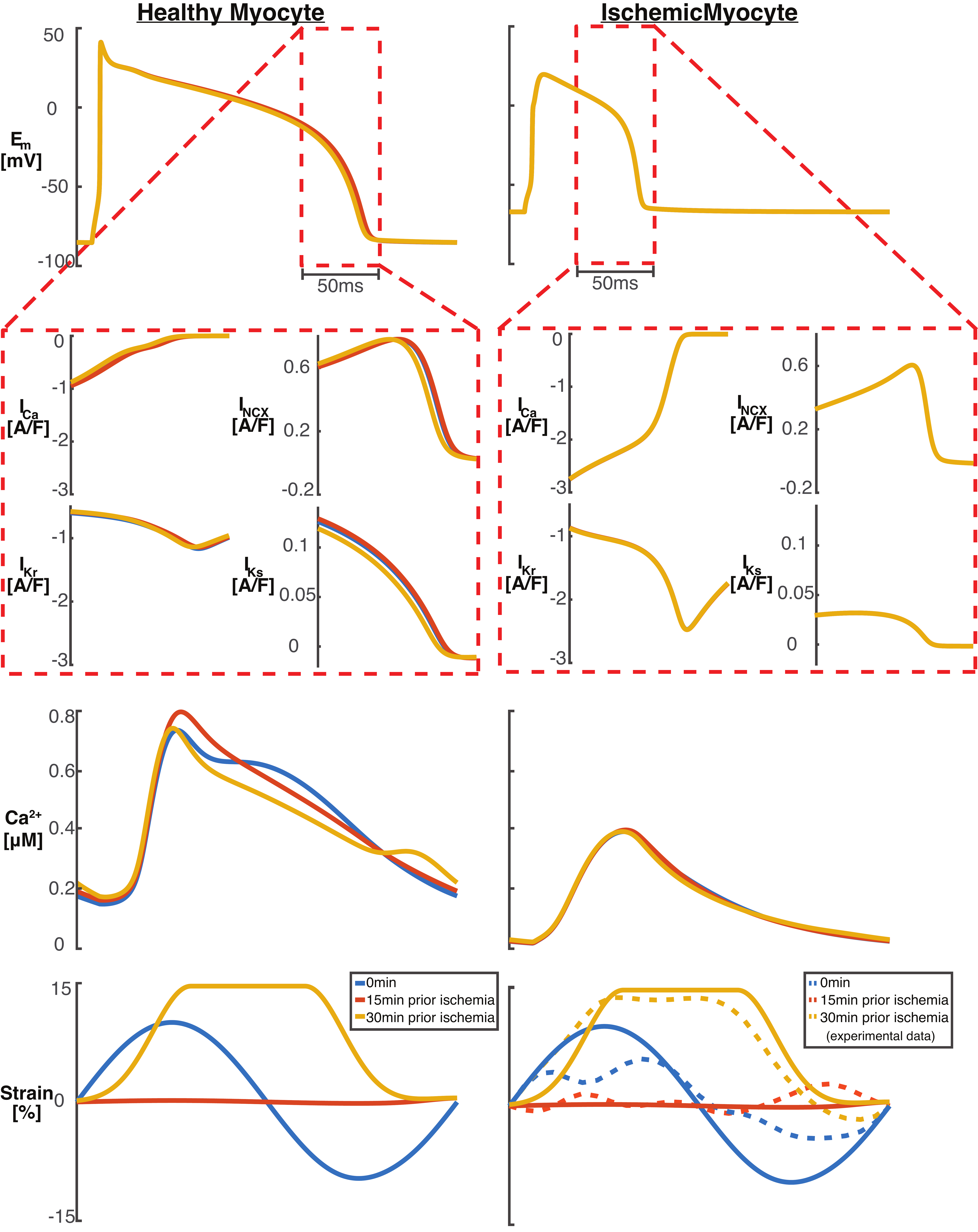}
\caption{Simulations of ventricular healthy myocytes and ischaemic myocytes with baseline stretch protocols (1), (2), and (3). These protocols reflect stretches observed experimentally before ischaemia [blue dotted line], after 15 min post-ischaemia initiation [red dotted line], and 30 min post-ischaemia initiation (yellow dotted line) \protect\cite{Baumeister2018} [dotted lines]. Our simulations indicate that stretch has a greater impact on CaT in healthy cells compared to ischaemic cells. However, there were no meaningful changes in action potential duration at 80\% repolarization (APD$_{\texttt{80}}$) noted in either the healthy or ischaemic myocytes.\label{figure:Figure2_second}}
\end{figure}

To investigate whether alterations in stretching due to the progression of ischaemia could be responsible for arrhythmias, we subjected our model to diverse stretching protocols based on experimental data \cite{Baumeister2018}. These protocols have been described in \cref{para:StretchProtocol} and are presented in \cref{figure:Figure2_second}).
Simulations using all three stretch protocols in all three myocyte models (healthy, border zone, and ischaemic) indicate that AP morphology remains unchanged (see \cref{figure:Figure2_second}).

In the study, we found that the healthy myocyte exhibited the highest sensitivity to different stretch protocols. This is attributed to the rapid change of sarcomere length at the end of the AP plateau phase with the stretch protocol (3), which induces the quickest adjustment of cytosolic calcium concentration due to the length-dependent changes in myofilament calcium buffering. As such, stretch protocol (3) yielded the most substantial impact on APD. 

\subsection{Simulations of mechanical and electrical heterogeneity do not affect the action potential} \label{section:ShannonSimulations}
To address mechanical heterogeneities observed in experiments, we investigated two alternative stretch protocols (stretch protocols (B) and (C) from \cref{para:StretchProtocol}). For both protocols, the modifications in CaT, as a result of length-dependent myofilament calcium sensitivity, did not substantially affect AP characteristics.

\begin{figure}[H]
\centering
\includegraphics[width=0.9\textwidth]{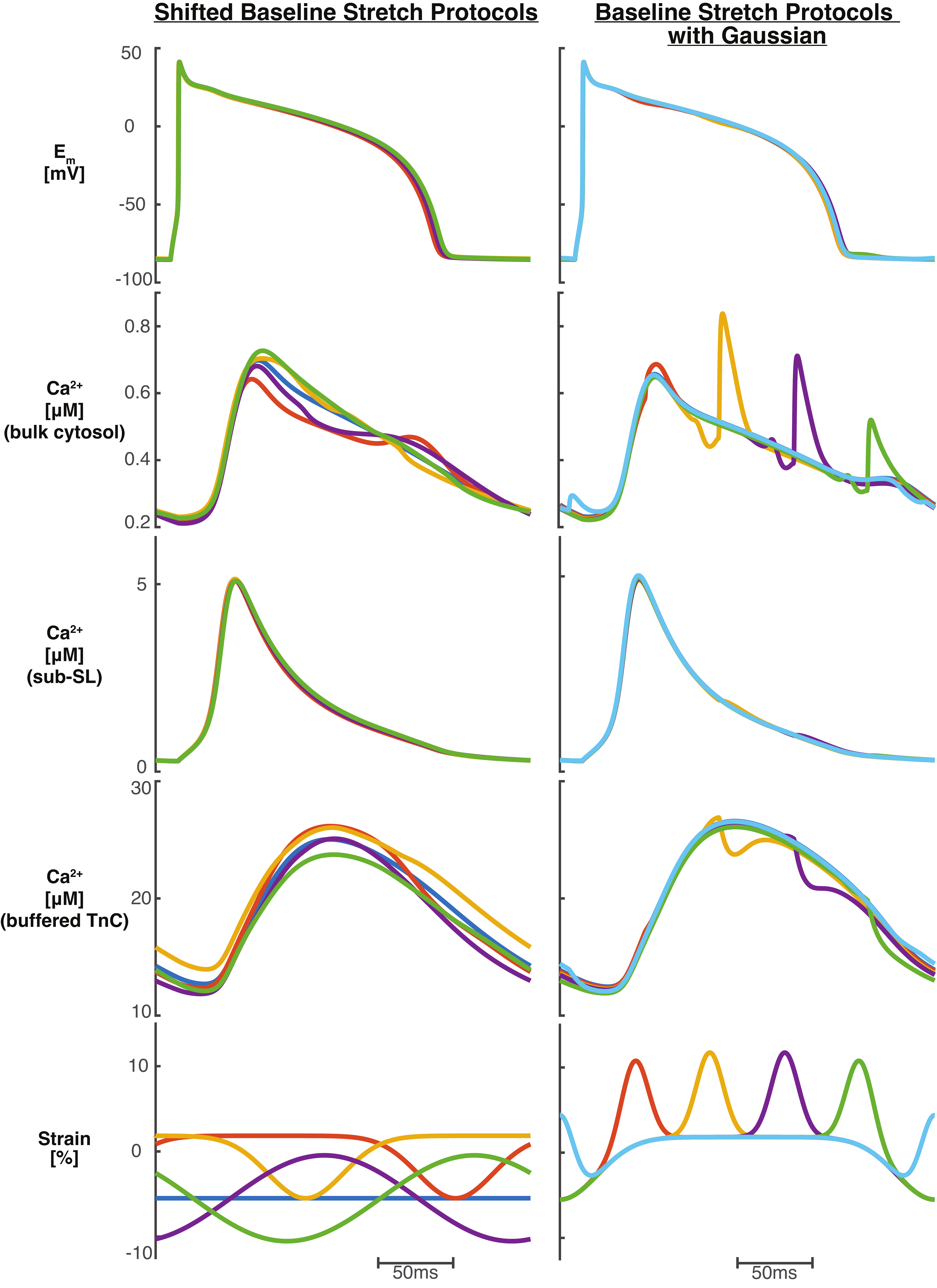}
\caption{Simulations of ventricular myocytes in healthy tissue with shifted baseline stretch protocols [left] and Gaussian-shaped stretches at 50 ms, 100 ms, 150 ms, and 200 ms [right]. While stretch protocols, that shifted baseline, mainly altered the amplitude of the CaT due to stretch-sensitive calcium release from the myofilaments, Gaussian-shaped stretches led to a meaningful change in CaT. Any changes in CaT did not affect the physiology of the APD$_{\texttt{80}}$.\label{figure:Figure3_second}}
\end{figure}

Including a rapid 10\% stretch at different timings on top of stretch protocol (C) within stretch protocol D from \cref{para:StretchProtocol} yields more pronounced alterations to the CaT, while the resulting APD$_{\texttt{80}}$ remains within physiological range. Nevertheless, the baseline stretch protocols, combined with pre-stretch and rapid stretches, can result in a total stretch of 40\%, which is not physiologically realistic.
Thus, the model did not demonstrate any considerable variations in APD, even in extreme conditions.

\begin{figure}[H]
\centering
\includegraphics[width=0.9\textwidth]{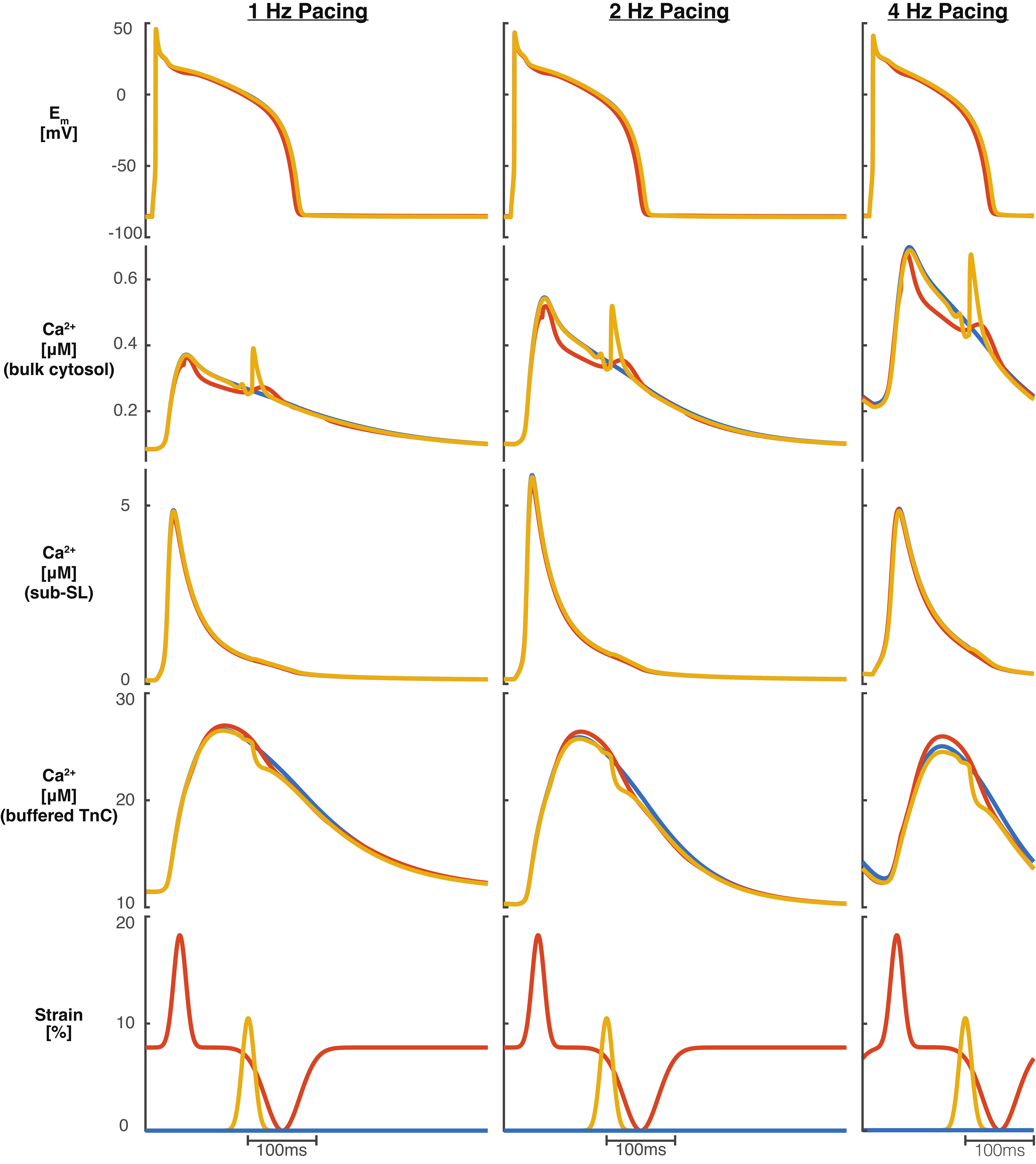}
\caption{Simulations at pacing rates of 1 Hz [left], 2 Hz [middle], and 4 Hz [right] for un-stretch [blue] and stretching protocols leading to a maximal [red] and minimal [yellow] change in APD$_{\texttt{80}}$ at 4 Hz pacing. With increasing pacing rates, the amplitude of the CaT increases and APD decreases. However, the magnitude of calcium released from the myofilaments due to stretch release scales with the changes of CaT amplitude. Therefore, with stretch at all pacing rates, the APD$_{\texttt{80}}$ is considered physiological.\label{figure:Pacing}}
\end{figure}

We examined the impact of the stretch protocol (C) described in \cref{para:StretchProtocol}, with was extended with an additional 15\% pre-stretch, on APD$_{\texttt{80}}$. Our findings, using the pre-stretched stretch protocol (C) with 30 perturbations, indicated that APD$_{\texttt{80}}$ ranged from 169.6 ms to 175.3 ms (red lines in \cref{figure:Figure3_second}; see \cref{table:APD80} for summarized findings). We noticed APD to be prolonged when Gaussian-shaped stretches were released at the end of the plateau phase. An increase in cytosolic calcium at the end of the CaT (purple line in \cref{figure:Figure3_second}; right). However, the constant stretching at the beginning of the AP and the release of the stretch towards the end of the plateau phase reduced the amplitude of the CaT, caused by a higher calcium affinity to TnC. This resulted in a shorter APD$_{\texttt{80}}$ (indicated by the red line in \cref{figure:Figure3_second}; left). In all simulations, the maximum deviation between APD and control was approximately 5 ms.

\begin{table}[H] 
\caption{Minimal and maximal APD$_{\texttt{80}}$ values for all combinations - shifted, pre-stretched, and additional rapid stretch - from the three different baseline stretch protocols. The most pronounced changes in APD$_{\texttt{80}}$ occurred for stretch protocol (2) (6.3 ms difference), whereas pre-stretch did not produce any meaningful changes in APD$_{\texttt{80}}$.\label{table:APD80}}
\newcolumntype{C}{>{\centering\arraybackslash}X}
\begin{tabularx}{\textwidth}{CCCC}
\toprule
\multicolumn{4}{c} {\textbf{Minimal and maximal APD$_{\texttt{80}}$ values}}\\
\midrule
& protocol 1 & protocol 2 & protocol 3\\
\midrule
\texttt{shift} & 173.7 ms & 175.1 ms & 171.3 ms\\
 & 175.9 ms & 175.1 ms & 174.8 ms\\
\hline
\texttt{+pre-stretch} & 173.7 ms & 170.6 ms & 171.8 ms\\
 & 176 ms & 170.6 ms & 171.6 ms\\
\hline
\texttt{+rapid stretch} & 172.6 ms & 174.5 ms & 169.6 ms \\
 & 174.6 ms & 175.3 ms & 173.7 ms\\
\hline
\texttt{+pre-stretch} & 172.6 ms & 169 ms & 171.7 ms \\
\texttt{+rapid stretch} & 176.2 ms & 169.2 ms & 172 ms\\
\bottomrule
\end{tabularx}
\end{table}

To investigate whether the pacing rate influences the impact of stretch on APD$_{\texttt{80}}$, we conducted several stretch protocols at pacing frequencies of  1, 2, and 4 Hz. Although faster pacing led to the shortening of APD and the augmentation of CaT amplitude in all simulated instances, stretch did not substantially alter APD$_{\texttt{80}}$. It is noteworthy that the CaT amplitude reduced with slower pacing rates leading to a decrease in calcium release from the myofilaments upon stretch release (see \cref{figure:Pacing}; left panel for 1 Hz pacing). However, the fraction of calcium released from the myofilaments compared to the CaT amplitude remains constant for all pacing rates. Therefore, APD$_{\texttt{80}}$ is similarly affected for all pacing rates, and APD$_{\texttt{80}}$ and is shortened by approximately 5 ms (see \cref{figure:Pacing}).

\subsection{Stretch during calcium overload conditions does not alter APD}
\begin{figure}[H]
\centering
\includegraphics[width=0.9\textwidth]{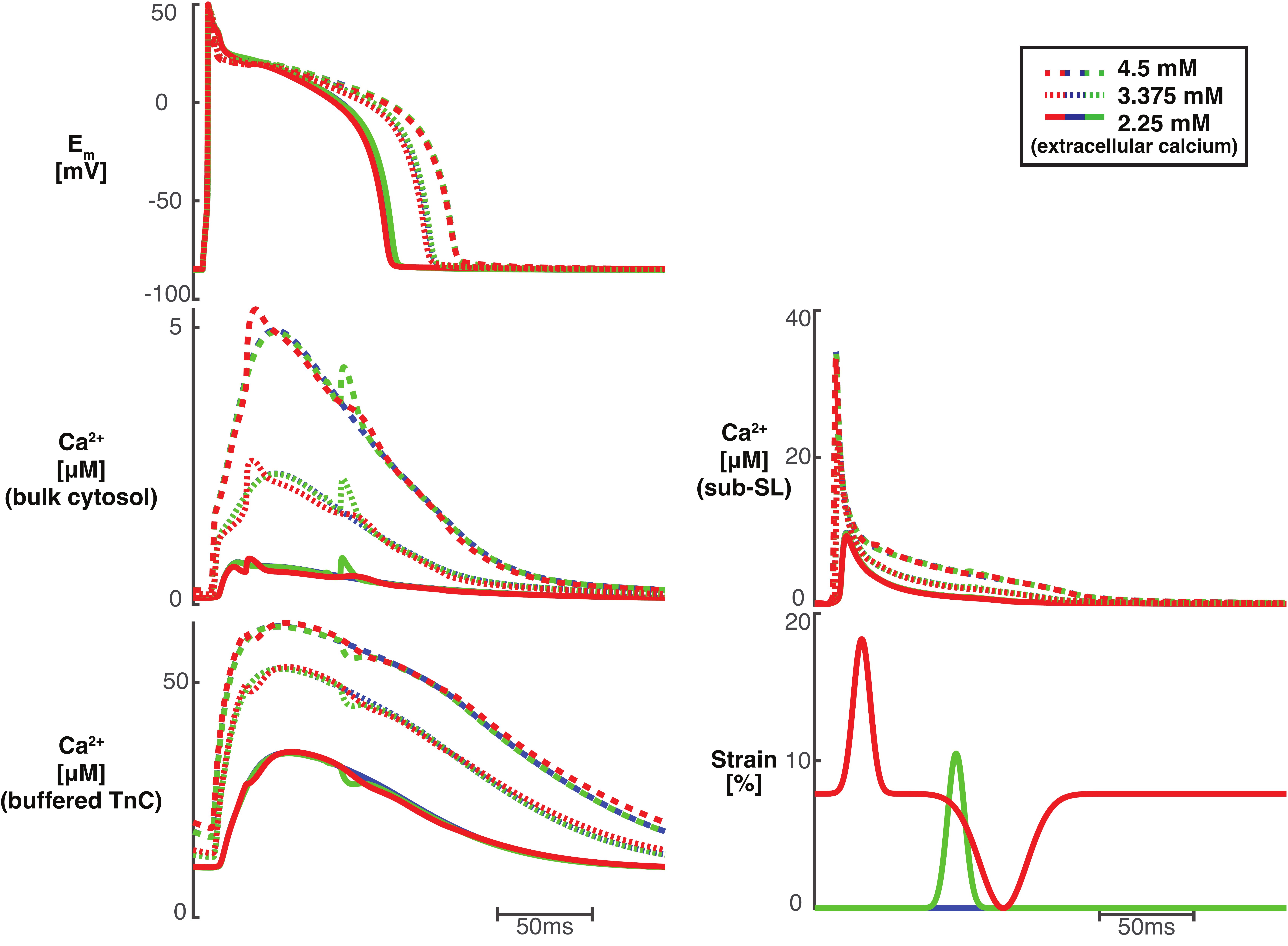}
\caption{Simulations at different extracellular calcium concentrations; $2.25$ mM (solid lines), $3.375$ mM (dotted lines), and $4.5$ mM (dashed lines) for three different stretch protocols. As extracellular calcium concentrations increase, the calcium concentration across all compartments increases, resulting in higher buffering on TnC. The total increase in cytosolic calcium and the total calcium release from the myofilaments remains similar for $3.375$ mM and $4.5$ mM compared to $2.25$ mM. Therefore, the most notable changes in APD$_{\texttt{80}}$ occur for $3.375$ mM extracellular calcium (dotted lines).
\label{figure:Figure6}}
\end{figure}  

We tested the model under calcium overload conditions to investigate whether the limited effect of MEC on APD was due to the relatively low amplitude of the baseline CaT. We elevated extracellular calcium by 25\% ($2.25$ mM, solid lines in \cref{figure:Figure6}), 87.5\% ($3.375$ mM, dotted lines in \cref{figure:Figure6}), and 150\% ($4.5$ mM, dashed lines in \cref{figure:Figure6}) and paced the model for 500 beats to reach steady state. Increased extracellular calcium levels under steady-state conditions lead to increased calcium concentrations  within the compartments (SR, sub-SL, cytosolic bulk, and dyad). In addition to higher calcium concentrations, we also applied rapid stretches at different timings together with stretch protocol (C). Interestingly, a 25\% rise in extracellular calcium concentration leads to a shortening of APD$_{\texttt{80}}$ (190.7 ms compared to 197.1 ms in control). Higher calcium concentrations have a lesser impact on APD (see \cref{figure:Figure6}) because of the elevated calcium concentration in the sub-SL, which is associated with gradient-based diffusion of calcium in the sub-SL from the cytosolic bulk. 

\subsection{Simulations of stretch with myofilaments in the sub-SL compartment}
\begin{figure}[H]
\centering
\includegraphics[width=0.9\textwidth]{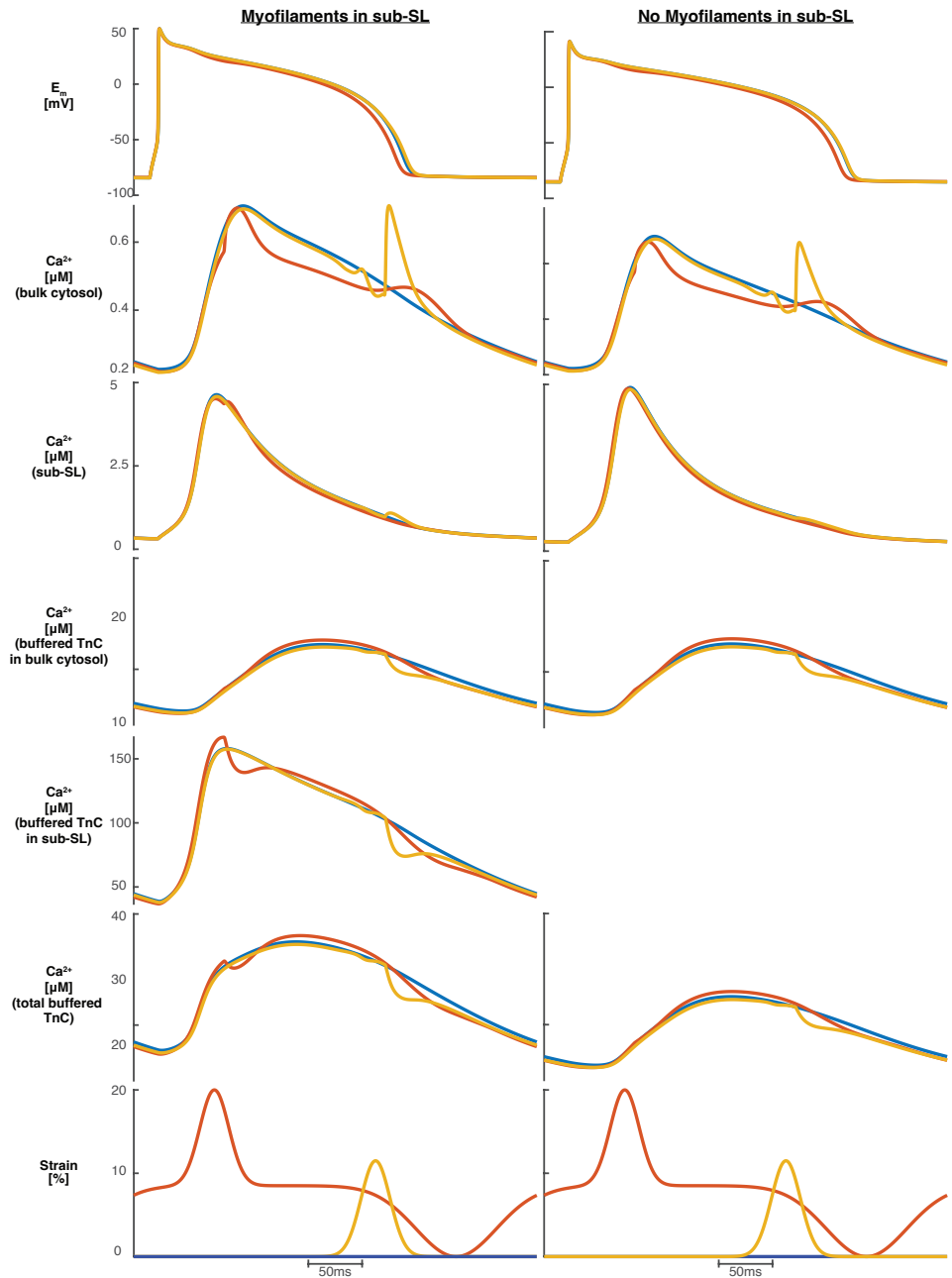}
\caption{Simulations for three different stretch protocols; no stretch [blue], shortened APD$_{\texttt{80}}$ [red], and prolonged APD$_{\texttt{80}}$ [green] for both control settings [right] and myofilaments in the sub-SL [left]. The model with myofilaments in the sub-SL exhibited a more pronounced effect on the total calcium buffering, and in turn, CaT. Despite this, the resulting APD remained in the physiological range.
\label{figure:Figure7}}
\end{figure}

The NCX is activated by an allosteric site on the exchanger located on the intracellular side of the membrane. Calcium dissociates from its activation site when the CaT falls to rest and the NCX deactivates. Consequently, the NCX switches off at a calcium concentration near resting cytosolic calcium levels, to prevent further calcium loss. This action ensures that the cell is not depleted of calcium \cite{Shannon2004}.

The sub-SL space incorporated in the employed EP model might hinder the effect of mechanically-induced alterations in CaT on APD$_{\texttt{80}}$. To address this limitation, we relocated 10\% of the myofilaments from the cytosolic bulk to the sub-SL compartment via the addition of TnC calcium buffering in the sub-SL. This method enabled a more direct impact of stretch effects on calcium buffering on membrane currents. 
However, the observed length-dependent changes in calcium concentration in the sub-SL did not meaningfully change APD$_{\texttt{80}}$. Thus, our examinations aimed to determine whether the calcium released from the myofilaments in the sub-SL diffuses to the cytosolic bulk before its binding to the activation sites of NCX. Due to the fast diffusion of calcium from the sub-SL to the cytosolic bulk, length-dependent modifications in myofilament calcium buffering were more pronounced in the cytosolic bulk as compared to previous simulations. Therefore, calcium released from the myofilaments in the sub-SL diffused into the cytosolic bulk without influencing the membrane potential (see \cref{figure:Figure7}). 

Faster diffusion between the sub-SL and cytosol ensures a faster equilibrium between calcium concentration in the sub-SL and cytosolic bulk towards the end of the AP plateau phase. Hence, alterations in cytosolic calcium concentration during equilibrium can affect sub-SL concentration. This novel formulation for accelerated calcium diffusion between sub-SL and cytosolic bulk promotes the effect of calcium release from the myofilaments on AP. Additionally, modified calcium diffusion alters calcium homeostasis by reducing sub-SL calcium concentration and elevating cytosolic and SR calcium concentration.

\section{Discussion}

This study demonstrates that mechanical perturbations during acute ischaemia can lead to experimentally observed calcium release events from the myofilaments, without producing meaningful changes in APD. In particular, AP morphology is the least susceptible in the ischaemic region, despite this region exhibiting the most pronounced vulnerability window (grey bars in \cref{figure:Figure1_second}). The ischaemic changes in EP probably lead to a reduction in CaT amplitude and APD.

Although changes in cytosolic calcium concentration induced by mechanical factors coincide with the occurrence of arrhythmia, our model is unable to replicate MEC effects as observed in experiments \cite{Boyden2001}. In particular, the strongly coupled mechanoelectrics model developed in this study includes a range of MEC features and mechanisms, it does not reproduce the EP changes observed experimentally during physiological strain and contraction dynamics. 

Here, two aspects of the results are examined: (a) potential restrictions of current computational models for cardiac cells, and (b) limitations of our model to depicting mechanically-induced calcium release from the myofilaments.

\subsection{Limitations of current myocyte models}
Experiments have shown that the initiation of calcium waves is in proximity to the border zones of mechanically-heterogeneous tissues, which can cause premature beats. These findings indicate that understanding the mechanisms behind arrhythmia requires a full comprehension of MEC, specifically triggered propagated contraction (TPC) \cite{terKeurs1998,terKeurs2006}. 

Although our investigation of this study was focused on acute ischaemia, our findings may have broader implications for state-of-the-art cardiac mechano-electrical models and their ability to accurately capture experimentally-observed MEC effects \cite{Timmermann2017}.
Mechanical perturbations in ventricular myocytes cause changes in CaT \cite{terKeurs1998,terKeurs2006}, demonstrating a close link between mechanical activity and calcium dynamics \cite{iribe2009axial,Prosser2011,Prosser2013}.

Previous computational investigations of MEC have primarily been focused on SAC (see exemplary \cite{Trayanova2011}). Our results suggest that while cellular EP models are unable to replicate the experimental results of calcium release from the myofilaments following stretch release, these modifications do not have a destabilising effect on the membrane potential. 
We propose two potential factors which may account for these outcomes: (a) the models were not parameterised using data that incorporates these behaviours, and/or (b) the fragility of cardiac tissue under pathological perturbations is incompatible with the needed robustness of computational models. 

The development of cardiac cell models has primarily focused on ECC to constrain channel kinetics, flux, current, and buffer dynamics. Experimental data on ECC has been used to develop EP models for different scales, levels of detail, and species \cite{Arevalo2016,Grandi2010,OHara2011,Oliveira2013,Shannon2004}. However, the development of those models did not consider MEC mechanisms. To our knowledge, no existing model used MEC data to limit electrophysiological kinetics and dynamics related to stress or strain, including accompanying calcium release from the myofilaments. While SAC were included in EP models at a later stage \cite{quinn2013combining}, previous research has verified that arrhythmogenic behaviour cannot be observed when scaled within the physiological range \cite{Trayanova2011}. 

The present study reveals the remarkable resilience and robustness of the tested models in response to diverse conditions and induced variations in calcium concentration, pacing, and calcium buffering on TnC.
The responsiveness and sensitivity of membrane-embedded channels to fluctuations in cytosolic calcium concentration or sub-SL was absent, even with the presence of a fractional relocation of the myofilaments to the sub-SL compartment. 
Therefore, varying ECC characteristics did not explain the diverse responses to alterations in cytosolic calcium concentration. 

This apparent ``over-stability" of the tested models may be a result of multiple factors. The heart is a highly complex system that is not yet completely understood, and computational models employ a mixture of deterministic and stochastic approaches. These models simplify and average experimental data, while integrating different heterogeneous experimental data from various scales, resolutions and modalities \cite{Kitano2002,kohl2010systems,quinn2011systems}. Additionally, the heart is a regulatory network compromising billions of cells (exact numbers depend on the species) with positive and negative feedback loops interacting in a complex and stochastic manner, yet structured. Also, the cardiac system must adapt dynamically to changes in pacing rate or possible perturbations like spontaneous calcium releases. Therefore, it needs to be robust at several levels. In fact, the heart needs to maintain its functionality against various natural and physiological perturbations while also flexibly adapting its mode of operation to achieve its specific function. This robustness arises from multiple control feedback loops, including ECC, MEC, concentration buffering, or signalling pathways, all of which incur a cost. Enhanced stability against specific perturbations may cause instability when encountering unexpected occurrences, such as cardiac disorders \cite{Kitano2002}. As per the highly optimised tolerance theory \cite{Carlson2002}, the systems' extreme fragility towards unknown perturbations can lead to catastrophic outcomes. Dynamic control systems, such as the heart, are optimized to balance the trade-offs between robustness and fragility, highlighting the importance of maintaining a finely tuned balance. To ensure stability and robustness in the presence of perturbations, a dynamic control system like the heart adapts its components to maintain normal (cardiac) function.
This specific attempt to account for perturbations can result in failure of the system \cite{Kitano2002,Kwon2008}. 

Despite the resiliency of the heart to physiological changes yet susceptibility to pathological changes, computational models are designed to withstand all types of perturbations to ensure model convergence and reproducibility. 

\subsection{Limitations of our representation of TPC}
Experiments indicate that TPC are spatially contractile events that cannot entirely be represented in a 0D model \cite{terKeurs1998,terKeurs2006}. Therefore, the models used in this study solely examine TPC triggers, and cannot probe or rule out whether the CaT in the model would not propagate or induce arrhythmias. Additionally, ischaemia is a constantly and dynamically changing process that proves challenging to model. All simulations of ischaemia were conducted under steady-state conditions, potentially falsifying the results. Similarly, the simulations of stretch events may be subject to dynamic changes over time, adjusting and varying upon the progression of the cardiac disorder. Consequently, the feedback to and from the myofilaments could affect the simulation outcomes. Finally, the mechanical kinetics model \cite{Rice2008} is not parameterized to CaT with low amplitudes as in the employed rabbit model (\cite{Shannon2004}). However, under calcium overload conditions, when the amplitude of the CaT reaches the range of the CaT in the mechanics model, the NCX switches into the reverse mode. This may lead to falsified simulation results. Furthermore, the (normalized) force traces for peak CaT concentrations lower than $0.85$ $\mu M$ are below 10\% of twitch forces, i.e. the transition from non-permissive to permissive states for the crossbridge cycle is almost certainly below 10\%. Hence, the likelihood of triggering calcium release events from the myofilaments through TPC is minimal.

\section{Conclusions}
The findings suggest that current ventricular myocyte models may not be capable of replicating MEC mechanisms, or that stretch-dependent alterations in myofilament calcium sensitivity induced by stretching and release may not impact APD. Computational myocyte models are generally not sensitive enough to reproduce mechanically-induced calcium release from the myofilaments. The observed variations in APD$_{\texttt{80}}$ were around 5 ms, which is meaningful, but still falls within normal physiological variation. To determine if these changes may cause pro-arrhythmic effects, tissue-level simulations are necessary, which was out of the scope of this study.

In this analysis, we examined the cellular mechanisms underlying changes in APD resulting from stretch in the presence of SAC and stretch-associated changes in myofilament calcium sensitivity in the context of acute ischaemia. The aim was to determine whether MEC mechanisms contribute to pro-arrhythmic effects during acute ischaemia. Using strongly coupled mechano-electrical models of rabbit ventricular myocytes, we demonstrated that the electrophysiological effects of both MEC mechanisms are responsible for alterations in intracellular calcium concentrations and are crucially reliant on the timing and velocity of stretch, as well as the calcium handling characteristics of the myocyte. However, due to the limited interaction between the cytosolic calcium and the AP, even substantial releases of calcium from the myofilaments do not result in meaningful changes in APD. Moreover, we observed that common pool models are inadequate in predicting myofilament release-dependent AP changes, because of the sub-SL compartment construction and the calcium diffusion parameterisation between sub-SL and cytosolic bulk. Finally, ischaemic tissue exhibits increased stability compared to healthy tissue, despite having a more pronounced susceptibility to trigger a depolarising current from NCX and subsequent EAD development within the vulnerability window. 



\vspace{6pt}

\abbreviations{Abbreviations}{
The following abbreviations are used in this manuscript:\\

\begin{tabular}{@{}ll}
ATP & adenosine triphosphate \\
AP & action potential\\
APD & action potential duration\\
APD$_{\texttt{80}}$ & action potential duration at 80\% repolarization\\
CaT & calcium transient \\
CaTD & calcium transient duration \\
CICR & calcium-induced calcium release \\
EAD & early after depolarization\\
ECC & excitation-contraction coupling\\
EP & electrophysiology \\
MEC & mechano-electrical coupling \\
NCX & sodium-calcium exchanger\\
ODE & ordinary differential equation\\
RyR & ryanodine receptor \\
SAC & stretch-activated current \\
SCD & sudden cardiac death\\
SL & sarcolemmal \\
SR & sarcoplasmic reticulum \\
TnC & troponin C \\
TPC & triggered propagated contraction
\end{tabular}
}




\begin{adjustwidth}{-\extralength}{0cm}

\reftitle{References}

\end{adjustwidth}
\end{document}